

\input phyzzx
\sequentialequations

\Pubnum{EPHOU-94-005}
\date{September 19, 1994}
\titlepage
\title{Is Hall Conductance in Hall Bar Geometry a Topological 
       Invariant?}
\author{{\rm K. Ishikawa, N. Maeda, and K. Tadaki}}

\address{Department of Physics, Hokkaido University, Sapporo 060, 
Japan}
\abstract
A deep connection between the Hall conductance in realistic 
situation and a topological invariant is pointed out based on 
von-Neumann lattice representation in which Landau level electrons have
minimum spatial extensions. We show that the Hall conductance has no 
finite size correction in quantum Hall regime, but  
a coefficient of induced Chern-Simons term in QED$_3$ has 
a small finite size correction, 
although both of them are similar topological invariant. 

\endpage

\doublespace

\hsize=469pt

\REF\a{K. v. Klitzing, G. Dorda, and M. Pepper, 
Phys. Rev. Lett. {\bf 45}, 494 (1980).}
\REF\b{R. B. Laughlin, 
Phys. Rev. {\bf B 23}, 5632 (1981).}
\REF\c{B. I. Halperin, 
Phys. Rev. Lett. {\bf 25}, 2185 (1982).}
\REF\d{D. J. Thouless, M. Kohmoto, M. P. Nightingale and M. den Nijs, 
Phys. Rev. Lett. {\bf 49}, 405 (1982); 
For a recent work, see Y. Hatsugai, Phys. Rev. Lett. 
{\bf 71}, 3697 (1993).}
\REF\e{A. Niu, D. J. Thouless, and Y. S. Wu, 
Phys. Rev. {\bf B 31}, 3372 (1985).}
\REF\f{K. Ishikawa and T. Matsuyama, 
Z. Phys. {\bf C 33}, 41 (1986); Nucl. Phys. {\bf B 280}, 
523 (1987).}
\REF\g{N. Imai, K. Ishikawa, T. Matsuyama, and I. Tanaka, 
Phys. Rev. {\bf B42}, 10610 (1990).}
\REF\h{R. Landauer, IBM J. Res. Dev. {\bf 1}, 223 (1957); 
M. B\"uttiker, Phys. Rev. {\bf B 38}, 9375 (1988).}
\REF\i{K. Ishikawa, 
Phys. Rev. Lett. {\bf 53}, 1615 (1984); 
Phys. Rev. {\bf D 31}, 1432 (1985).}
\REF\j{R. Jackiw, 
Phys. Rev. {\bf D 29}, 2375 (1984).}
\REF\l{A. Niemi and G. Semenoff, 
Phys. Rev. Lett. {\bf 51}, 2088 (1983); 
A. Redlich, 
Phys. Rev. Lett. {\bf 52}, 18 (1984).}
\REF\m{K. Ishikawa, in Proceedings of the 3rd International 
Symposium of Fundation of Quantum Mechanics, the Physical 
Society of Japan, 1990, p.70.}
\REF\n{T. Matsuyama, Prog. Theor. Phys. {\bf 77}, 711 (1987).}
\REF\o{H. So, Prog. Theor. Phys. {\bf 74}, 585 (1985).}
\REF\p{S. Coleman and B. Hill, 
Phys. Lett. {\bf 159B}, 184 (1985); 
Y. C. Kao and M. Suzuki, Phys. Rev. {\bf D 29}, 2137 (1985).}
\REF\q{O. Abe and K. Ishikawa, in {\it Rationale of Being: 
Festschrift in honor of Gyo Takeda}, edited by K.Ishikawa 
{\it et al}. (World Scientific, Singapore, 1986), p.137.}
\REF\r{A. M. Perelomov, Theor. Mat. Fiz. {\bf 6}, 213 (1971); 
V. Bargmann {\it et al}., Rep. Math. Phys. {\bf 2}, 221 (1971).}
\REF\s{K. Ishikawa, N. Maeda, and K. Tadaki, 
Hokkaido University preprint, EPHOU-94-002 (unpublished).}

Since the discovery of the quantum Hall effect,$^{\a}$ 
topological origin 
of quantization of the Hall conductance has been discussed. 
It was initiated by Laughlin$^{\b}$ 
who gave a simple argument for showing 
a quantization in a torus system. He used the fact that the torus 
system is invariant under a finite gauge transformation, which 
can be considered as adding a unit of flux to the inside of the 
torus. The response of the electronic system gives an average Hall 
conductance over a unit of flux because the flux is not infinitesimal. 
Thus the quantization of the average Hall conductance was shown. 
The idea was extended by Halperin$^{\c}$ 
and the role of edge was identified. 

An idea of using topological invariant was introduced later by 
TKNdN.$^{\d}$ 
They have identified the Hall conductance in tight binding 
model as first Chern class of the fibre bundle. 
An argument was extended further later,$^{\e}$ in which average over 
boundary condition is taken. 

For the topological invariant to be defined, the corresponding space 
must be compact. In the above cases, invariance of the system under 
gauge transformation, or the model itself ensure the compactness. 
They are not applied in real cases, however, because the electron 
system is described neither on the torus nor by the tight binding 
model, and average over flux or boundary condition is not taken. 
Instead, the system is described by continuum space model in a finite 
area with boundary. The conductance is defined by a derivative also. 

A new idea of topological invariant in momentum space was introduced 
by one of the present authors$^{\f,\g}$ 
and others later, and low energy theorem 
was given. Recently, the quantization is also shown by using 
Landauer formula$^{\h}$ 
for one dimensional edge current under the assumption 
that only the edge states are current carrying states. 

The purpose of the present work is 
to unify all pictures and 
to find out if the Hall conductance 
is a topological invariant or it has a finite size correction in 
realistic systems, where there are zero-energy one-dimensional edge 
states. Our conclusion is that the Hall conductance in the realistic 
situation is equivalent to the topological winding number in the 
momentum space and the quantization is thus exact in the quantum Hall 
regime. The role of edge states is also clarified. 
A similar topological invariant of 
three-dimensional quantum electrodynamics(QED$_3$), a coefficient of 
induced Chern-Simons term,
has a small finite size corrections, of order 
$e^{-mL}$, where $L$ is the system size and $m$ is the typical mass 
scale. 

The roles of the magnetic field are essential in making not only the 
space to be compact but also the finite size effect to disappear. 

We describe an infinite system first, a finite system without boundary 
second, and finite system with boundary at the end.

(1) {\it Infinite plane without boundary} --- 
Before discussing the quantum Hall system, let us discuss 
QED$_3$ which resembles the quantum Hall system in many respects. 
$^{\i-\l}$
The slope of the current correlation function of momentum, 
$\pi_{\mu\nu}(q)$, at the origin is given by 
$$
{1\over3!}\epsilon^{\mu\nu\rho}{\partial\over\partial q_\rho}
\pi_{\mu\nu}(q)\biggr\vert_{q=0}=
{1\over3!}\int {d^3 p\over(2\pi)^3}\epsilon^{\mu\nu\rho}
\Tr[{\partial S^{-1}(p)\over\partial p_{\rho}}S(p)
{\partial S^{-1}(p)\over\partial p_{\mu}}S(p)
{\partial S^{-1}(p)\over\partial p_{\nu}}S(p)],
\eqno\eq
$$
$$
S^{-1}(p)=p_\mu\gamma^\mu-m, \ 
\gamma^0=\sigma^3,\ \gamma^1=
i\sigma^1,\ \gamma^2=i\sigma^2
.
$$
Obviously the slope has a peculiar meaning as a winding number of the 
mapping defined by the propagator $S(p)$,$^{\m}$ 
if the momentum is defined 
on the compact space. Arguments concerning its topological nature
$^{\n,\o}$ 
and the non-renormalization theorem$^{\p,\q}$ 
have been given in the literature. 
It should be noted that the non-trivial topology can be defined 
even in 
infinite planar system and a torus configuration is not necessary 
in QED$_3$. 
For non-renormalization theorem to be satisfied, current conservation 
and Ward-Takahashi identity are important. 

It is convenient to use a representation in which base functions are 
localized spatially in each Landau level in order to derive 
Ward-Takahashi identity and related exact low energy theorem in 
quantum Hall system. 
Coherent state von-Neumann lattice representation$^{\g}$ 
in centre variables 
is such representation that has minimum spatial extensions allowed 
from commutation relations and its usefulness has been shown in 
previous works in showing the low energy theorem and others. 
We use this representation in the present work, too, and give 
low energy theorems of finite systems. 

In our coherent state von-Neumann lattice representation,$^{\g}$ 
the kinetic term and the conserved electromagnetic current are written 
as, 
$$
\eqalign{
\int d^2x&\psi^\dagger(x){(\vec p+e\vec A)^2\over2m}\psi(x)
=\sum_{l,m,n}E_l b_l(m,n) a_l(m,n),\cr
&\partial_xA_y-\partial_yA_x=B,\ E_l={eB\over m}(l+{1\over2}),}
\eqno\eq
$$
$$
\eqalign{
&j_\mu(x)=\cr
{eB\over m}&\sum_{l_i,m_i,n_i}b_{l_1}(m_1,n_1) a_{l_2}(m_2,n_2)
\int{d^2k\over(2\pi)^2}\langle \tilde R_{m_1,n_1}\vert e^{
i\vec k\cdot\vec X}\vert R_{m_2,n_2}\rangle 
(f_{l_1}\xi_\mu e^{i(k_x \xi+k_y \eta)}f_{l_2})e^{-i\vec k\cdot\vec x}
,} 
\eqno\eq
$$
where $\xi_\mu=(1,-\eta,\xi),\ 
eB\xi=p_y+eA_y,\ eB\eta=-(p_x+eA_x),\ X=x-\xi,\ Y=y-\eta$, and 
electron field is expressed by quantized operators 
$a_l(m,n)$ and $b_l(m,n)$, and c-number functions of relative 
coordinates and centre coordinates:
$$
\eqalign{
\psi&=\sum a_l(m,n)f_l(\xi,\eta)\ket{R_{m,n}},\cr
\psi^\dagger&=\sum b_l(m,n)f_l(\xi,\eta)\langle\tilde R_{m,n}\vert,}
\eqno\eq
$$
$$
\eqalign{&
{e^2B^2\over 2m}(\xi^2+\eta^2)f_l(\xi,\eta)=E_l 
f_l(\xi,\eta),\cr
&(X+iY)\ket{R_{m,n}}=a(m+in)\ket{R_{m,n}},\cr
&\langle \tilde R_{m_1,n_1}\vert R_{m_2,n_2}\rangle=
\delta_{m_1,m_2}\delta_{n_1,n_2}-1/\sum_{m,n},\cr
}
\eqno\eq
$$
where $R_{m,n}=(ma,na)$ and $m$, $n$ are integers.
The coherent states in centre coordinates become minimum set of 
complete set$^{\r}$ if lattice spacing 
$a$ is equal to $\sqrt{2\pi/
eB}$, which we use hereafter. The dual basis $\langle\tilde R_{m,n}
\vert$ 
is obtained by a linear combination of $\langle R_{m,n}\vert$ 
with suitable 
coefficients, as was given in Refs.\g\ and \s.
One-particle states, such as localized states around short range 
impurity, edge states, and extended states under periodic potentials, 
were studied in this representation before.$^{\s}$ 

The electron field operators, $a_l(m,n)$ and $b_l(m,n)$, have 
lattice coordinates and Landau level indices. 
Hence the electron propagator and the vertex part become matrices 
which have Landau level indices and momentum defined on the torus. 
The momentum space is thus compact in the spatial directions. 

The matrix relation between the propagator and the vertex part 
is derived from the current conservation and commutation relations 
and becomes to
$$
\eqalign{
q^\mu \Gamma_\mu(p_1,p_2)&=S^{-1}(p_1)R(p_1,p_2)-L(p_1,p_2)S^{-1}
(p_2),\cr
R_{l_1,l_2}(p_1,p_2)&=\delta_{l_1,l_2}+iq_j[d_j(p_2)\delta_{l_1,l_2}
+\bar d_{j,l_1,l_2}],\cr
L_{l_1,l_2}(p_1,p_2)&=\delta_{l_1,l_2}+iq_j[d'_j(p_2)\delta_{l_1,l_2}
+\bar d'_{j,l_1,l_2}],}
\eqno\eq
$$
where Landau level dependent terms in the above matrices satisfy,
$$
[\bar d_x,\bar d_y]=[\bar d'_x,\bar d'_y]=-i/eB.
\eqno\eq
$$
The vertex part of the same momenta thus satisfies a modified 
Ward-Takahashi identity, 
$$
\eqalign{
\Gamma_\mu(p,p)&={\partial\over\partial p_\mu}S^{-1}(p)+S^{-1}(p)
D_\mu-D'_\mu S^{-1}(p),\cr
D_0&=D'_0=0,\ D_i=i(d_i+\bar d_i),\ D'_i=i(d'_i+\bar d'_i).\cr}
\eqno\eq
$$
By a transformation of the basis with suitable 
momentum dependent unitary matrices, 
$$
\eqalign{
\tilde S(p)&=V(\vec p)S(p)U(\vec p),\cr
\tilde \Gamma_\mu(p_1,p_2)&=U^{-1}(\vec p_1)\Gamma_\mu(
p_1,p_2)V^{-1}(\vec p_2),}
\eqno\eq
$$
$$
U(\vec p){\partial\over\partial p_i}U^{-1}(\vec p)=-D'_i,\ 
({\partial\over\partial p_i}V^{-1}(\vec p))V(\vec p)=D_i,
\eqno\eq
$$
the second term and third term in 
right hand side of Eq.(8) are absorbed and the relation 
becomes to the standard form, 
$$
\tilde \Gamma_\mu(p,p)={\partial\over\partial p_\mu}\tilde S^{-1}(p).
\eqno\eq
$$
The Hall conductance is the slope of the current correlation function, 
$\pi_{\mu\nu}(q)$, at the origin and is expressed, by using Eq.(11), 
as a topologically invariant form, 
$$
\eqalign{
\sigma_{xy}&={e^2\over3!}\epsilon^{\mu\nu\rho}{\partial\over\partial q_
\rho}\pi_{\mu\nu}(q)\biggr\vert_{q=0}={e^2\over2\pi}N_w,\cr
N_w&=
{1\over24\pi^2}\int d^3p\epsilon^{\mu\nu\rho}
\Tr[{\partial \tilde S^{-1}(p)\over\partial p_{\rho}}\tilde S(p)
{\partial \tilde S^{-1}(p)\over\partial p_{\mu}}\tilde S(p)
{\partial \tilde S^{-1}(p)\over\partial p_{\nu}}\tilde S(p)].}
\eqno\eq
$$
We have the momentum independent propagator from Eq.(2),
$$
S^{(0)}(p)_{l_1,l_2}=(p_0-E_{l_1})^{-1}\delta_{l_1,l_2}
\eqno\eq
$$
in the free system without disorder, boundary, and interactions. 
The $\sigma_{xy}$ thus calculated depends on $E_F$, as is 
shown in Fig.1. 
The integrand is independent from the spatial component of the momenta 
from Eqs.(7) and (9). 
This property becomes important when we discuss finite system. 

In the presence of short range disorders and interactions,  
the above formula is valid, as well, because the 
current conservation and the commutation relations are kept intact. 
Moreover the value of $\sigma_{xy}$ is not modified by these effects 
in the localized state region and in the energy gap region. 
More details are given in Ref.\g. 

The Hall conductance in the infinite planar system is expressed with a 
topologically invariant formula and is quantized in the integer filling 
region and neighboring localized state region and is not modified by 
interactions. These low energy theorems concerning the Hall conductance 
have been given in the present representation. The matrix form of the 
Ward-Takahashi identity between the vertex part and the propagator 
plays the essential roles.

(2) {\it Finite plane without boundary} --- 
In the formula of the Hall conductance, Eq.(12), integration regions 
of the momentum is from $-\pi/a$ to $\pi/a$. Now suppose the 
configuration space to be a torus of length $L$. The momentum then 
becomes discrete and the integration is replaced with a summation over 
discrete momentum. 
We show in the following that the topological invariant, Eq.(12), is 
unchanged and has the exactly same value but another topological 
invariant, Eq.(1), changes the value. 
We assume $L/2a$ being integer and study the $L$ dependence. 
The momentum then becomes $2\pi n/L$ with an integer $n$ in the range 
from $-L/2a$ to $L/2a$. 

The Hall conductance in a torus geometry in $y$ direction is given by 
$$
\sigma_{xy}={e^2\over2\pi}N'_w,
\eqno\eq
$$
where $N'_w$ is obtained by replacing $p_y$ integration of $N_w$, 
Eq.(12), 
with a summation over discrete $p_y$. 
Since the integrand is independent from $p_y$, the above $N'_w$ is the 
same as the previous $N_w$, Eq.(12), and is an exact integer under 
integer Hall effect condition. 
The absence of finite size correction in the Hall conductance is 
partly due to the special form of the modified 
propagator, Eq.(9), and commutation relation, Eq.(7), which have 
an origin in the magnetic field. 

The finite size correction exists in the topological invariant of 
QED$_3$, Eq.(1). It changes the value under the replacement of 
momentum integration to summation over discrete momentum and 
becomes to $\coth(mL/2)$. It has a finite size correction of order 
$e^{-mL}$.

(3) {\it Finite plane with boundary\ (edge states)} --- 
We discuss the quantum Hall effect in a realistic situation, in which 
electrons are confined in a finite area by a potential barrier. 
We concentrate to find out the effect of the potential 
barrier here. The single-particle Hamiltonian is composed of three 
diagonal terms, 
$$
\eqalign{
H_0&=\int d^2x
\psi^\dagger(x)[{(\vec p+e\vec A)^2\over2m}+V(x)]\psi(x)\cr
&=\sum_{\rm inside}E_l b_l(X)a_l(X)+
\sum_{\rm outside}(E_l+V_0) b_l(X)a_l(X)+
\sum_{\rm boundary}(E_l+\Delta E^{(\alpha)})b^{(\alpha)}a^{(\alpha)},}
\eqno\eq
$$
$$
0\leq\Delta E^{(\alpha)}\leq V_0,\quad
V(x)=\cases{0,&inside,\cr V_0,&outside.\cr}
$$
Eigenstates from the first and second term have the energy of Landau 
level, 
or the combined energy of Landau level and the potential energy, 
respectively. The third term gives the continuum energy band in an 
energy range between $E_l$ and $E_l+V_0$ and eigenstates are extended 
along the boundary. They correspond to the edge states, 
which may modify the previous results. 

It is convenient to write the above Hamiltonian, 
in order to find the effect of edge states, as 
$$
H_0
=\sum_{x<x_0+\Delta}E_l b_l(X)a_l(X)+
\sum_{x\geq x_0+\Delta}(E_l+V_0) b_l(X)a_l(X)+
\sum \Delta E_{l,l'}(X,X')b_l(X)a_{l'}(X'),
\eqno\eq
$$
where the spatial region in the first includes inside and edge region, 
and $\Delta E$ of the last term vanishes when the coordinates 
$\vec X$ or $\vec X'$ is located away from the baundary. 
Hence we are able to treat the last term perturbatively. 
The second term gives no effect except the virtual effect, if 
the potential height $V_0$ is much larger than the chemical potential. 
We treat the first term in the same manner as before and 
include the last term to it perturbatively. 

Due to the last term in Eq.(16), the propagator, the vertex, and the 
current correlation function are not invariant under the 
translation. Even so, the translational invariant expression is valid. 
Green's functions have invariant terms and non-invariant terms. 
The invariant terms contribute to the Hall conductance and 
non-invariant terms do not contribute to the conductance. 
The invariant term, $\pi^{(0)}_{\mu\nu}(q)$, and non-invariant terms, 
$\pi^{(1)}_{\mu\nu}(q_1,q_2;l_1)$, $\pi^{(2)}_{\mu\nu}(q_1,q_2;l_1,l_2)
\cdots,$ of current correlation functions are given in 
Fig.2, where dashed lines show the perturbative edge terms which carry 
momentum $l_i$. $\pi^{(1)}_{\mu\nu}(q_1,q_2;l_1)$ satisfies 
\nextline
$q_1^\mu\pi^{(1)}_{\mu\nu}(q_1,q_2;l_1)=\pi^{(1)}_{\mu\nu}(q_1,q_2;l_1)
q_2^\nu=0$. If $q_1$ and $q_2$ are independent arbitrary momentum, 
then $\pi^{(1)}_{\mu\nu}(q_1,q_1;0)$ has no linear term in $q_1$ 
if this is obtained smoothly by a limit $q_2\rightarrow q_1$, 
by Coleman-Hill argument.$^{\p}$ Other general term 
$\pi^{(n)}_{\mu\nu}(q_1,q_2;l_1,\cdots l_n)$ satisfies the same relation 
and the same properties for general $l_i$. 
The amplitude of present specific case, in which momentum along one 
direction is conserved, is obtained as a limit of general cases and 
should satisfies the same properties, as far as there is no singularity 
and the limit is taken smoothly. We find that 
each term is non-singular and the combined edge term is also 
non-singular. 
The most serious part is from edge state and is computed from 
one-dimensional chiral mode. We find that it is non-singular. 
Consequently only the one loop diagram, $\pi^{(0)}_{\mu\nu}(q)$, has 
linear term in $q$ and contributes to $\sigma_{xy}$. 
This diagram was studied in the previous part and was shown to give 
exactly quantized value in the quantum Hall regime. 
In other words, both of the edge states and the bulk states contribute 
to the quantized value of $\sigma_{xy}$. 
The value is stable even though there are zero-energy one-dimensional 
edge states. In Landauer formula, the zero-energy states are 
important.$^{\h}$ 
Since the zero-energy states are one-dimensional in the quantum Hall 
regime, the $\sigma_{xy}$ is quantized from this view point, too. 
Short range impurities and an electron interactions do not modify 
our conclusions at all, because all ingredients we used are still 
valid. 

Ultra-violet divergence is generated from virtual effect and 
renormalization is made. The results still hold, due to the same 
reasons. 

As a summary, we find that the small finite size correction exists in 
QED$_3$, but the finite size correction 
does not exist in the normal quantum Hall region. 
It is the magnetic field that gives these phenomena to occur. 

The present work is partially supported by the special Grant-in-Aid 
for promotion of Education and Science in Hokkaido University 
Provided by the Ministry of Education, Science and Culture, 
a Grant-in-Aid for general Scientific Research(03640256), and the 
Scientific Research on Priority Area(04231101), the Ministry of 
Education, Science and Culture, Japan.

\FIG\fia{Hall conductance, from Eq.(12), has a step like dependence 
on the Fermi energy. The value is stable and has no correction 
from impurities and interactions in the gap region and localized region
(shaded region). }

\FIG\fib{Feynman diagram of current-current correlation function is 
shown. The perturbative edge terms carry momentum and are shown by 
dashed lines. The wavy lines stand for the currents. 
Since the edge terms carry momentum, two momenta, $q_1$ and $q_2$, 
are different generally. }

\endpage
\refout
\endpage
\figout
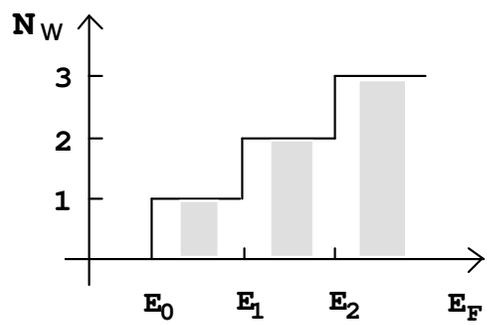
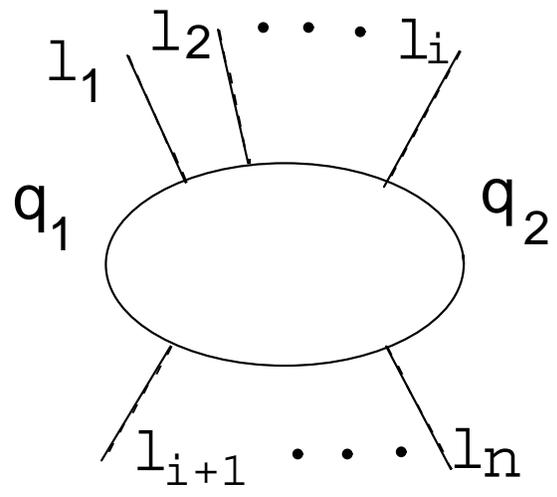
\end